\documentclass[a4paper]{article}

\usepackage{INTERSPEECH2022}
\usepackage{multirow}
\usepackage{mathtools}
\usepackage{subcaption}
\usepackage{graphicx}
\usepackage{url}

\title{Exploiting Hidden Representations from a DNN-based Speech Recogniser for Speech Intelligibility Prediction in Hearing-impaired Listeners}
\name{Zehai Tu, Ning Ma, Jon Barker}
\address{University of Sheffield, Department of Computer Science, Sheffield, UK}
\email{\{ztu3, n.ma, j.p.barker\}@sheffield.ac.uk}

\begin{document}

\maketitle
\begin{abstract}
An accurate objective speech intelligibility prediction algorithms is of great interest for many applications such as speech enhancement for hearing aids. Most algorithms measures the signal-to-noise ratios or correlations between the acoustic features of clean reference signals and degraded signals. However, these hand-picked acoustic features are usually not explicitly correlated with recognition. Meanwhile, deep neural network (DNN) based automatic speech recogniser (ASR) is approaching human performance in some speech recognition tasks. This work leverages the hidden representations from DNN-based ASR as features for speech intelligibility prediction in hearing-impaired listeners. The experiments based on a hearing aid intelligibility database show that the proposed method could make better prediction than a widely used short-time objective intelligibility (STOI) based binaural measure.

\end{abstract}
\noindent\textbf{Index Terms}: Intelligibility prediction, objective measures, deep neural networks, hearing-impaired listeners, hearing aids

\section{Introduction}
Accurate objective speech intelligibility measurement plays an important role in the development of speech enhancement, e.g. hearing aids, because subjective listening experiments can be time-consuming and expensive~\cite{falk2015objective}. Since the 2010s, objective intelligibility prediction algorithms, including STOI~\cite{taal2011algorithm}, sEPSM~\cite{jorgensen2011predicting}, and HASPI~\cite{kates2014hearing}, have achieved success by comparing the acoustic features of reference and degraded speech signals. However, these algorithms have been rarely evaluated on the degraded speech enhanced by data-driven DNN-based models, which have made significant progress in speech enhancement~\cite{wang2018supervised}.

The speech recognition performance of recent DNN-based ASR models is approaching that of humans, and they have also shown
similar patterns in speech recognition results~\cite{barker2007modelling, schadler2015matrix, fontan2017automatic, arai2020predicting}. Therefore, it has been of interest to use DNN-based ASR for intelligibility prediction. Meanwhile, DNNs are naturally good feature extractors. Compared to the acoustic features proposed in the aforementioned intelligibility prediction algorithms, hidden representations of DNN-based ASR are optimised to directly correlate with recognition. 

In this work, we exploit the hidden representations from one of the state-of-the-art end-to-end ASR models for intelligibility prediction on the first round Clarity Prediction Challenge (CPC1)~\cite{barker2022the}, which includes a large number of binaural speech signals simulated in complex noisy environments and then enhanced by complex hearing-aid models. The results show that the similarities between ASR hidden representations of reference and processed signals could be better than the conventional acoustic features at intelligibility prediction. The results also show that using ASR hidden representations could be a better method for intelligibility prediction than simply using ASR recognition results.

This paper is organised as follows. Section~\ref{sec:relatedwork} reviews related objective intelligibility prediction measures. Section~\ref{sec:method} presents the ASR model used in this work and the method for similarity computation from hidden representations. Section~\ref{sec:setup} shows the experimental setup including database and evaluation methods. The results and detailed analyses are presented and discussed in Section~\ref{sec:results}. Section~\ref{sec:conclusions} concludes the work and proposes future plans.

\section{Related Work}
\label{sec:relatedwork}
Early work on intelligibility prediction for speech in additive noise is based on measured signal-to-noise ratio (SNR) in frequency bands or at modulation frequencies, e.g., the articulation index~\cite{kryter1962methods}, the speech intelligibility index~\cite{ansi1997methods} and the speech transmission index~\cite{steeneken1980physical}. However, these methods are not suitable in scenes involving (strong) non-stationary noises and nonlinear processing~\cite{graetzer2021intelligibility}. A more recent SNR-based algorithm is sEPSM, which calculates the ratio between envelope powers of processed signals and residual noise. 
Similarly, GEDI~\cite{yamamoto2017predicting} predicts intelligibility based on the signal-to-distortion ratio in the envelopes output from a gammachirp auditory filterbank. More recent methods have been correlation-based, with STOI being the most widely used example \cite{andersen2018refinement}. STOI and its variants are based on the cross-correlation between the temporal envelopes output from a one third octave filterbank of the reference and processed signals. However, 
they produce poor estimates for noisy speech enhanced by Wiener filtering~\cite{yamamoto2017predicting} or DNN-based enhancement~\cite{gelderblom2017subjective}. Additionally, HASPI incorporates a hearing loss model, and combines the measures of auditory coherence and cepstral correlation between the reference signal and the processed signal degraded by hearing loss. Most of these methods use perceptually motivated acoustic features, while not directly tuned to be correlated with speech recognition.

Previous works using ASR models for intelligibility prediction usually leverage outputs of ASR models, and most of them are evaluated on small vocabulary tasks. For example, the predicted words are used to compute the recognition correctness in~\cite{schadler2015matrix, fontan2017automatic}. Alternatively, phoneme posterior probabilities from the ASR outputs are used to estimate the ASR word error rates~\cite{martinez2021prediction} or directly predict speech recognition thresholds~\cite{rossbach2021non} for matrix sentence tests. Several word posterior related measures are explored in~\cite{karbasi2020non}. For unlimited vocabulary size, phone accuracy and phone posterior are used in~\cite{arai2019predicting} and~\cite{arai2020predicting}. To the best of our knowledge, ours is the first work to use ASR hidden representations for intelligibility prediction. The proposed method could take better advantage of the reference signals, and work particularly well in some situations, e.g., when the processed and reference signals are very different, while the ASR still makes a correct \textit{guess}.

\section{Method}
\label{sec:method}
In this section we describe how we leverage hidden representations from an ASR model for intelligibility prediction. ASR based on transformer architectures~\cite{vaswani2017attention} has achieved great success recently and is used in this work. The predicted intelligibility is defined as the similarity between hidden representations from reference signals $x$ and processed signals $\hat{x}$. In addition, performances of hidden representations at different levels within the ASR are investigated. Although binaural signals produced by hearing aids are studied, single-channel ASR is used for hidden representation extraction, that is, two hidden representations for left and right channels are extracted given a binaural signal. All four hidden representations from the left, right channels of reference and processed signals $H^l$, $H^r$, $\hat{H}^l$, $\hat{H}^r$ are used to compute the similarity. 

\subsection{ASR model}
\begin{figure}[t]
  \centering
  \includegraphics[width=0.7\linewidth]{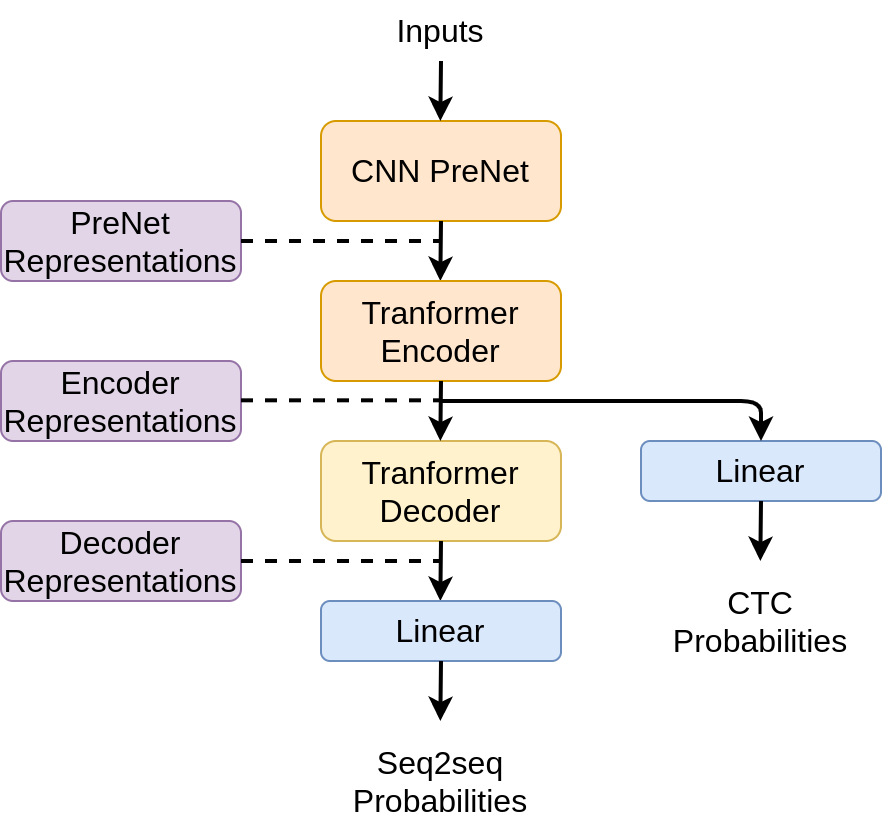}
  \caption{ASR architecture and hidden representations at three levels.}
  \label{fig:asr}
\end{figure}

Figure~\ref{fig:asr} shows the architecture of the transformer-based ASR model used in this work. It consists of a convolutional neural network (CNN) based PreNet, a transformer-based encoder, and a transformer-based decoder. The PreNet is a stack of convolutional layers for better understanding global context~\cite{han2020contextnet}. Both the encoder and decoder are composed of a number of transformer blocks. Each encoder transformer block consists of a multi-head self-attention sub-layer and a position-wise fully connected feed-forward sub-layer. A residual connection and layer normalisation~\cite{ba2016layer} are applied to both sub-layers. In the multi-head attention sub-layer, the input features are firstly mapped to query $Q$, key $K$ with embedding length $d_{k}$, and value $V$ with embedding length $d_{v}$, and the attention mechanism is computed as:
\begin{equation}
    \textit{Attention}\left(Q, K, V \right) = \textit{Softmax}\left(\frac{Q K^{T}}{\sqrt{d_{k}}} \right)V.
\end{equation}
The projection and attention mechanism are run in parallel multiple times, and the concatenation of attention outputs is then multiplied by a linear projection matrix. Compared to the encoder transformer block, an extra multi-head attention sub-layer is inserted to perform the attention mechanism on the encoder output features. In addition, a positional mask is used in the decoder multi-head self-attention sub-layer to enforce that only the known previous decoded outputs are dependent.

The ASR model is optimised with the joint CTC-attention mechanism~\cite{kim2017joint}, i.e., a combination of Connectionist Temporal Classification (CTC)~\cite{graves2006connectionist} and attention-based sequence-to-sequence (seq2seq)~\cite{chorowski2015attention}. The CTC leverages repeatable intermediate label representation and a special blank label for ASR decoding, and the loss function can be expressed as:
\begin{equation}
    \mathcal{L}_\mathnormal{CTC} = - \log \left( \sum_{\pi \in \beta^{-1}(l)} \prod_{m=1}^{M} P(z^{m}_{\pi_{m}}) \right),
\end{equation}
where $\beta$ is a function that removes repeated intermediate and blank labels, $\pi_{m}$ is the intermediate and blank label sequence, $P(z^{m}_{\pi_{m}})$ is the probability of $\pi_{m}$ at time $m$, and $l$ is the target label sequence. The seq2seq loss function is the sum of divergences between the ground truth label $z_{u}$ and predicted token $\hat{z}_{u}$ at $u$-th position in the transcript sequence:
\begin{equation}
    \mathcal{L}_\mathnormal{seq2seq} = \sum_{u} P(z_{u})(\log P(z_{u}) - \log P(\hat{z}_{u})).
\end{equation}
The overall loss function for ASR optimisation is:
\begin{equation}
    \mathcal{L} = \lambda\mathcal{L}_\mathnormal{CTC} + (1 - \lambda)\mathcal{L}_\mathnormal {seq2seq},
\end{equation}
where $\lambda$ is a predefined weighting coefficient.

\subsection{Hidden representations}
Three hidden representations shown in Figure~\ref{fig:asr} are studied in this work, including outputs of the CNN PreNet ${H}^{pre} \in \mathcal{R}^{T^{pre} \times d^{pre}}$, outputs of the transformer encoder ${H}^{enc} \in \mathcal{R}^{T^{enc} \times d^{enc}}$, and outputs of the transformer decoder ${H}^{dec} \in \mathcal{R}^{T^{dec} \times d^{dec}}$. The PreNet representations ${H}^{pre}$ are viewed as low-level acoustic features. Meanwhile, the encoder representations ${h}^{enc}$ can be viewed as high-level acoustic representations, as ASR models using CTC decoding does not learn a language model, and CTC output intermediate labels are independent from each other. In contrast, the seq2seq decoder is usually considered as an internal language model~\cite{meng2021internal}. Therefore, the decoder representations ${H}^{dec}$ are viewed as hidden representations with learnt language knowledge.

\subsection{Similarity computation}
The cosine similarity is used in this work as it is naturally well scaled. Given two hidden representations at a single time step, $h, \hat{h} \in \mathcal{R}^{d}$, the cosine similarity is computed as:
$\cos(h, \hat{h}) = \frac{h \cdot \hat{h}}{\Vert h  \Vert \Vert \hat{h} \Vert}$,
where $\Vert \cdot \Vert$ is the $L2$ norm. For PreNet and encoder representations, the reference and processed representations of each time step are matched. The similarity at each time step $\rho_{t}$ for the binaural reference and processed representations is computed from the pair of representations at this time step, i.e., $\rho_t = \cos(h_t, \hat{h}_t)$.   The overall similarity between the binaural reference and processed representations is computed from the four pairs of hidden representations at each time step, i.e., $\{h^l, \hat{h}^l\}$, $\{h^l, \hat{h}^r\}$, $\{h^r, \hat{h}^l\}$, $\{h^r, \hat{h}^r\}$:
\begin{equation}
    \mathnormal{sim}(H^{bi}, \hat{H}^{bi}) = \frac{1}{T}\displaystyle\sum_{t=1}^{T} \max\left\{\rho_{t}^{ll}, \rho_{t}^{lr}, \rho_{t}^{rl}, \rho_{t}^{rr} \right\}.
\end{equation}

For decoder representations, the left and right representations of the reference and processed signals could have variable time steps, i.e., $T^l$, $T^r$, $\hat{T}^l$, $\hat{T}^r$ could be different. For each pair of sequences of decoder representations $\{H, \hat{H} \}$, the fast dynamic time warping algorithm~\cite{salvador2007toward} is applied to find the warp path. The overall similarity of the warped pair is computed as:
\begin{equation}
    \mathnormal{sim}(H_{w}, \hat{H}_{w}) =  \frac{1}{T_{w}}\displaystyle\sum_{t=1}^{T_{w}} \cos (H_{w}(t), \hat{H}_{w}(t)),
\end{equation}
where $H_{w}$ and $\hat{H}_{w}$ are the warped representations, $T_{w}$ is the total time steps after warping. The overall binaural similarity is then computed as:
\begin{equation}
\begin{split}
    \mathnormal{sim}(H^{bi}, \hat{H}^{bi}) &= \max \bigg\{\mathnormal{sim}(H^l_{w}, \hat{H}^l_{w}), \mathnormal{sim}(H^l_{w}, \hat{H}^r_{w}), \\  & \mathnormal{sim}(H^r_{w}, \hat{H}^l_{w}), \mathnormal{sim}(H^r_{w}, \hat{H}^r_{w}) \bigg\}.    
\end{split}
\end{equation}



\section{Experimental Setup}
\label{sec:setup}
\subsection{Database}
CPC1 provides a large number of processed binaural speech signals by machine learning hearing-aid systems and the corresponding responses from hearing impaired listeners. Each signal represents a simulated mixture of a target speech and an interfering noise within a simulated cuboid-shaped living room, enhanced by a hearing-aid system given the audiogram (i.e., pure-tone measure of hearing thresholds at different frequencies) of a listener. Both the binaural processed signals and the corresponding anechoic reference signals are provided. The ground truth intelligibility is presented as the listener word correctness scores (WCS). A total of 6 speakers, 10 hearing aid systems and 27 listeners are included. The CPC1 includes two tracks: (1) \textit{closed-set}, that is the listeners and systems in the evaluation set are overlapped with those in the training data; (2) \textit{open-set}, that is the systems or listeners in the evaluation set are not included in the training data. For full details see~\cite{barker2022the}. For both tracks, the scenes in the training data are split into 70\% and 30\% as a training set and a development set, and the results on the evaluation set are reported.

\subsection{ASR model}
The SpeechBrain~\cite{ravanelli2021speechbrain} LibriSpeech transformer ASR recipe is used in this work. 80-channel log mel-filter bank coefficients are used as input with a 25\,ms window with a stride of 10\,ms. The PreNet consists of three 2D convolutional layers, and the encoder and the decoder consists of twelve and six transformer blocks, respectively. The weighting coefficient $\lambda$ is set 0.3 for training, and 0.4 for decoding. The dimensions at one time step for PreNet, encoder, and decoder hidden representations are 10240, 768, and 768, respectively.

\subsection{Evaluation}
The baseline intelligibility predictor includes the Cambridge MSBG hearing loss model~\cite{baer1993effects, baer1994effects, moore1993simulation, stone1999tolerable} and MBSTOI~\cite{andersen2018refinement}. The MSBG hearing loss model simulates hearing abilities given a listener's audiogram, and the MBSTOI is a refined binaural version of STOI. In addition, the correlations between WCS of listeners and of ASR models are also reported. 

Three performance evaluation measures, including root mean square error\,(RMSE), normalised cross-correlation coefficient\,(NCC), and Kendall’s Tau coefficient\,(KT), are exploited as the evaluation metrics. As the first two metrics could be invalid for non-linear correlations, a logistic function $f(x) = 1 / [1 + \exp(ax + b)]$ is applied to the predicted intelligibility to examine the monotonic relation, following the conventions of previous works, including \cite{taal2011algorithm, andersen2018refinement}. For the proposed method, the parameters $a$ and $b$ are optimised on the development set. For the baseline system, the parameters are optimised on all the CPC1 training data, as described in~\cite{barker2022the}.

\section{Experiments and Results}
\label{sec:results}

\begin{table}[t]
\centering
\caption{Evaluation results on the CPC1 of various methods.}
\resizebox{0.8\linewidth}{!}{
\begin{tabular}{l|c|c|c}
\toprule
& RMSE $\downarrow$ & NCC $\uparrow$ & KT $\uparrow$  \\\midrule
\multicolumn{4}{l}{Closed-set} \\ \midrule
Baseline & 0.285 & 0.621 & 0.398 \\
ASR WCS & 0.250 & 0.729 & \textbf{0.523} \\
PreNet representations & 0.347 & 0.299 & 0.182 \\
Encoder representations & 0.237 & 0.758 & 0.487 \\
Decoder representations & \textbf{0.231} & \textbf{0.773} & 0.498 \\ \midrule
\multicolumn{4}{l}{Open-set} \\ \midrule
Baseline & 0.365 & 0.529 & 0.391 \\
ASR WCS & 0.250 & 0.723 & \textbf{0.534} \\
PreNet representations & 0.356 & 0.254 & 0.136 \\
Encoder representations & 0.241 & 0.751 & \textbf{0.534} \\
Decoder representations & \textbf{0.235} & \textbf{0.763} & 0.530 \\
\bottomrule
\end{tabular}
}
\label{tab:representations}
\end{table}

\begin{figure}[t]
  \centering
    \begin{subfigure}{\linewidth}
        \includegraphics[width=\linewidth]{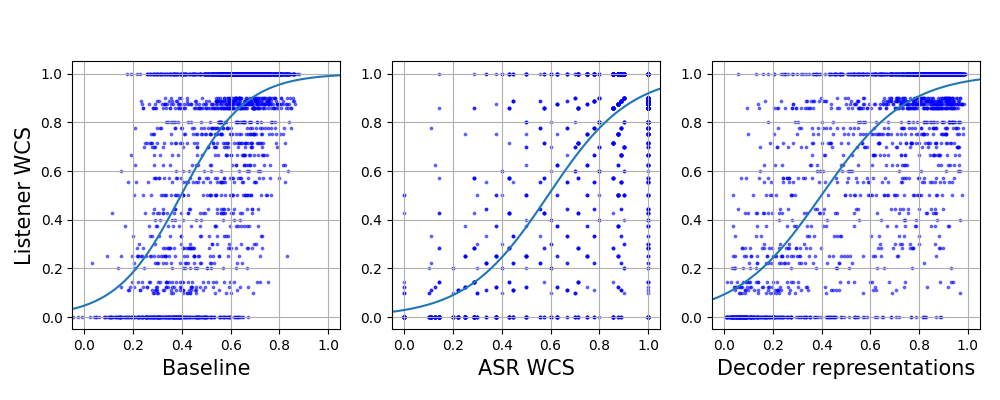}
        \caption{Closed-set}
    \end{subfigure}
    \begin{subfigure}{\linewidth}
        \includegraphics[width=\linewidth]{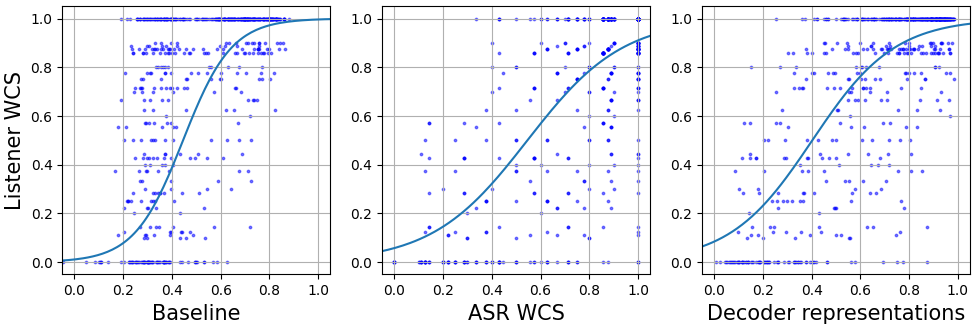}
        \caption{Open-set}
    \end{subfigure}
  \caption{Listener WCS versus predicted intelligibility by various methods and the corresponding logistic mapping functions.}
  \label{fig:cpc1}
\vspace{-3mm}
\end{figure}

\subsection{Hidden representations}
The training of the default ASR model starts from the pretrained model on the LibriSpeech\footnote{huggingface.co/speechbrain/asr-transformer-transformerlm-\protect\\librispeech} (LS). Therefore, it has a strong knowledge on clean speech. Furthermore, it is optimised with LS \textit{train-clean-100} set added with noises from the training set in the first round Clarity Enhancement Challenge~\cite{graetzer2021clarity} (CLS) for ten epochs. Finally, the ASR model is optimised on CPC1 training set for another ten epochs. In addition, the MSBG hearing loss model is used to process the signals when training and testing on CPC1. The correlations between the listener WCS and the baseline (MSBG+MBSTOI) prediction, ASR WCS, predicted intelligibility with different ASR hidden representations, are shown in Table~\ref{tab:representations}. Figure~\ref{fig:cpc1} also shows the listener WCS against the predicted intelligibility by the baseline, ASR WCS, the decoder representations, and their corresponding logistic mapping functions. Both the results of the \textit{closed-set} and \textit{open-set} indicate that the similarity between the reference and processed high-level hidden representations could outperform the baseline and ASR WCS at intelligibility prediction in terms of RMSE and NCC. The ASR WCS predictions are advantageous with regard to KT because WCS is discrete, i.e., in which case \textit{tied} pairs are more likely to appear. Between the two high-level hidden representations, the decoder ones including language model knowledge are better than the encoder ones which represent high-level acoustic features in terms of RMSE and NCC, while the KT scores are close. The following experiments and analyses are conducted on the \textit{closed-set}.

\begin{table}[t]
\centering
\caption{Evaluation results on the \textit{closed-set} of decoder representations from different ASR models.}
\resizebox{0.8\linewidth}{!}{
\begin{tabular}{c|l|c|c|c}
\toprule
MSBG & Training data & RMSE $\downarrow$ & NCC $\uparrow$ & KT $\uparrow$  \\\midrule
\multirow{4}{*}{with} & LS & 0.264 & 0.692 & 0.449 \\ 
& LS+CLS & 0.243 & 0.746 & 0.464 \\ 
& LS+CPC1 & 0.233 & 0.768 & \textbf{0.503} \\
& LS+CLS+CPC1 & \textbf{0.231} & \textbf{0.773} & 0.498 \\
\midrule
w/o & LS+CLS+CPC1 & 0.234 & 0.767 & 0.476\\
\bottomrule
\end{tabular}
}
\label{tab:data}
\end{table}

\begin{table}[t]
\centering
\caption{Listener- and system-wise evaluation results on the \textit{closed-set} of predicted intelligibility.}
\resizebox{0.8\linewidth}{!}{
\begin{tabular}{l|c|c|c}
\toprule
& RMSE $\downarrow$ & NCC $\uparrow$ & KT $\uparrow$  \\\midrule
\multicolumn{4}{l}{Listener-wise} \\ \midrule
Baseline &  0.078 & 0.414 & 0.311 \\
Decoder representations & 0.078 & 0.419 & 0.407\\ \midrule
\multicolumn{4}{l}{System-wise} \\ \midrule
Baseline & 0.147 & 0.798 & 0.244 \\
Decoder representations& 0.048 & 0.982 & 0.644\\
\bottomrule
\end{tabular}
}
\label{tab:avescores}
\vspace{-4mm}
\end{table}

\subsection{Data mismatch}
For the purpose of investigating the influence of data mismatch (i.e., different distribution of training and evaluation data) on ASR models for intelligibility prediction, four different ASR models with different training data knowledge (LS, LS+CLS, LS+CPC1, LS+CLS+CPC1) are probed. The MSBG model is used for all models as preprocessing for hearing loss simulation. ASRs trained on CLS can be considered to have knowledge of noisy speech, and those trained on CPC1 can be considered to have knowledge of processed noisy speech by hearing-aid systems. The correlations between the predicted intelligibility with decoder representations and the ground truth WCS are shown in Table~\ref{tab:data}. The results show that the ASR models trained with CPC1 training data (LS+CPC1, LS+CLS+CPC1) could make optimal predictions, while the latter one is slightly better in terms of RMSE and NCC because it has knowledge of noisy speech. Meanwhile, the ASR models with no knowledge of CPC1 data (LS, LS+CLS) could also achieve competitive results. It is worth noting that the ASR model trained only on clean LS signals could still outperform the baseline system.

\subsection{MSBG hearing loss model}
The influence of the MSBG hearing loss model is also investigated.  The intelligibility prediction results of ASR models trained on LS+CLS+CPC1 with and without the MSBG model for hearing loss simulation are also shown in Table~\ref{tab:data}. The results indicate that the MSBG hearing loss model could offer a slight advantage on intelligibility prediction for the ASR hidden representations.

\subsection{Listener- and system-wise correlation}
\begin{figure}[thb]
  \centering
  \begin{subfigure}{\linewidth}
  \includegraphics[width=\linewidth]{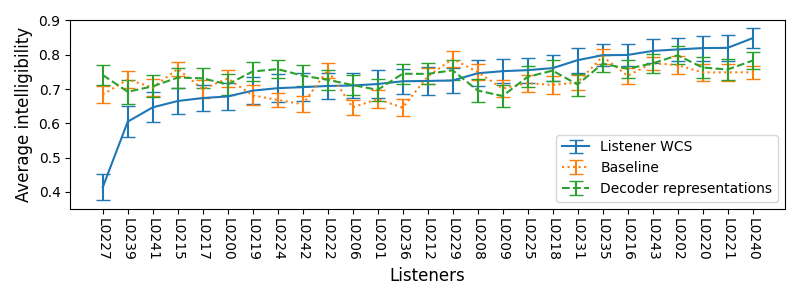}
  \caption{Listener-wise}
  \label{fig:listener}
  \end{subfigure}
  
  \begin{subfigure}{\linewidth}
  \centering
  \includegraphics[width=0.75\linewidth]{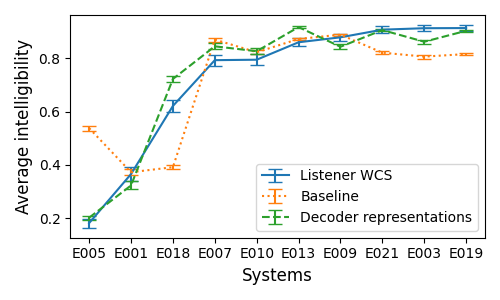}
  \caption{System-wise}
  \label{fig:system}
  \end{subfigure}
  \caption{Listener- and system-wise average intelligibility with standard errors on the \textit{closed-set}.}
\end{figure}

The results of the listening experiments provided by CPC1 can be noisy, because of the not strictly constrained speech materials, the large size vocabulary, etc. Therefore, both the listener WCS and the predicted intelligibility scores are averaged on listeners and hearing-aid systems for more conclusive analysis. The average listener WCS, the predicted intelligibility from the baseline, and the proposed decoder representation similarity from the ASR trained on LS+CLS+CPC1 with the MSBG hearing loss model on different listeners and hearing-aid systems with their corresponding error bars are shown in Figure~\ref{fig:listener} and Figure~\ref{fig:system}. The listener- and system-wise evaluation results on the \textit{closed-set} are measured and shown in Table~\ref{tab:avescores}. The intelligibility predicted from the ASR decoder representations could gain a slight advantage over the baseline on listener-wise intelligibility scores. For system-wise intelligibility, the proposed method is much more correlated with the listener WCS than the baseline.

\section{Conclusions}
\label{sec:conclusions}

This paper has proposed a novel method for speech intelligibility prediction by leveraging DNN-based ASR hidden representations. The similarity of the hidden representations from an ASR neural network is measured between a clean reference signal and the corresponding processed signal which is used for intelligibility prediction. The experimental results on the recent CPC1 database, which provides machine learning hearing aid processed binaural signals, have shown that the proposed method can outperform other intrusive methods such as MBSTOI, the refined binaural variant of STOI. Future work will focus on expanding the evaluation of the proposed method to different databases, such as~\cite{barker2007modelling}, with more baseline methods, including STOI and HASPI. 

\bibliographystyle{IEEEtran}
\bibliography{mybib}
\end{document}